\documentclass[12pt]{iopart}

\usepackage[english]{babel}
\usepackage[dvips]{graphicx}
\usepackage{amstext,amssymb}
\usepackage[comma,square,numbers,sort]{natbib}
\usepackage{array}

\def\21{SU(2) $\otimes$ U(1) }
\newcommand{\newblock}{}
\newcommand{\eqref}[1]{(\ref{#1})}

\newcommand{\dof}  {d.o.f.}
\newcommand{\eVq}  {\text{eV}^2}

\newcommand{\Dmq}  {\Delta m^2}
\newcommand{\Dms}  {\Delta m^2_{21}}
\newcommand{\Dma}  {\Delta m^2_{31}}

\begin{document}


\title{Three-flavour neutrino oscillation update}

\author{Thomas Schwetz\dag, Mariam
  T{\'o}rtola\ddag\ and Jos{\'e} W.~F.~Valle\S}

\address{\dag\ Theory Division, Physics Departement, CERN
  CH--1211 Geneva 23, Switzerland}

\address{\ddag\  Departamento de F\'\i sica and CFTP, 
Instituto Superior T\'ecnico\\
  Av. Rovisco Pais 1, 1049-001 Lisboa, Portugal}

\address{\S\ AHEP Group, Instituto de F\'{\i}sica Corpuscular --
  C.S.I.C./Universitat de Val{\`e}ncia, \\
  Edificio Institutos de Paterna, Apt 22085, E--46071 Valencia, Spain}

\ead{schwetz@cern.ch, mariam@cftp.ist.utl.pt, valle@ific.uv.es}

\begin{abstract}
  We review the present status of three-flavour neutrino oscillations,
  taking into account the latest available neutrino oscillation data
  presented at the {\em Neutrino 2008\/} Conference. This includes the
  data released this summer by the MINOS collaboration, the data of
  the neutral current counter phase of the SNO solar neutrino
  experiment, as well as the latest KamLAND and Borexino data. We give
  the updated determinations of the leading 'solar' and 'atmospheric'
  oscillation parameters. We find from global data that the mixing
  angle $\theta_{13}$ is consistent with zero within $0.9\sigma$ and
  we derive an upper bound of $\sin^2\theta_{13} \leq 0.035 \,
  (0.056)$ at 90\%~CL (3$\sigma$).
\end{abstract}

\maketitle

\section{Introduction}
\label{sec:introduction}

Thanks to the synergy amongst a variety of experiments involving
solar and atmospheric neutrinos, as well as man-made neutrinos at
nuclear power plants and accelerators~\cite{exp-talks-nu08} neutrino
physics has undergone a revolution over the last decade or so.
The long-sought-for phenomenon of neutrino oscillations has been
finally established, demonstrating that neutrino flavor states
$(\nu_e,\nu_\mu,\nu_\tau)$ are indeed quantum superpositions of states
$(\nu_1,\nu_2,\nu_3)$ with definite masses $m_i$~\cite{Amsler:2008zz}.
The simplest unitary form of the lepton mixing matrix relating flavor
and mass eigenstate neutrinos is given in terms of three mixing angles
$(\theta_{12},\theta_{13},\theta_{23})$ and three CP-violating phases,
only one of which affects (conventional) neutrino
oscillations~\cite{schechter:1980gr}.
Here we consider only the effect of the mixing angles in current
oscillation experiments, the sensitivity to CP violation effects
remains an open challenge for future
experiments~\cite{Bandyopadhyay:2007kx,Nunokawa:2007qh}.
Together with the mass splitting parameters $\Dms \equiv m^2_2-m^2_1$
and $\Dma \equiv m^2_3- m^2_1$ the angles $\theta_{12}, \theta_{23}$
are rather well determined by the oscillation data. In contrast, so
far only upper bounds can be placed upon $\theta_{13}$, mainly
following from the null results of the short-baseline CHOOZ reactor
experiment \cite{Apollonio:2002gd} with some effect also from solar
and KamLAND data, especially at low $\Dma$
values~\cite{Maltoni:2003da}.

Here we present an update of the three-flavour oscillation analyses of
Refs.~\cite{Maltoni:2003da} and \cite{Maltoni:2004ei}.  This new
analysis includes the data released this summer by the MINOS
collaboration~\cite{Adamson:2008zt}, the data from the neutral current
counter phase of the SNO experiment (SNO-NCD)~\cite{Aharmim:2008kc},
the latest KamLAND~\cite{:2008ee} and
Borexino~\cite{Collaboration:2008mt} data, as well as the results of a
recent re-analysis of the Gallex/GNO solar neutrino data presented at
the Neutrino 2008 conference~\cite{gallex-nu08:07}.
In Section~\ref{sec:lead-solar-atmosph} we discuss the status of the
parameters relevant for the leading oscillation modes in solar and
atmospheric neutrinos. In Section~\ref{sec:th13} we present the
updated limits on $\theta_{13}$ and discuss the recent claims for
possible hints in favour of a non-zero value made in
Refs.~\cite{Balantekin:2008zm,Escamilla:2008vq,Fogli:2008jx}.
We summarize in Section~\ref{sec:summary}.

\section{The leading 'solar' and 'atmospheric' oscillation parameters}
\label{sec:lead-solar-atmosph}

Let us first discuss the status of the solar parameters $\theta_{12}$
and $\Dms$.
The latest data release from the KamLAND reactor
experiment~\cite{:2008ee} has increased the exposure almost fourfold
over previous results~\cite{Araki:2004mb} to $2.44 \times 10^{32}$
proton$\cdot$yr due to longer lifetime and an enlarged fiducial
volume, corresponding to a total exposure of 2881~ton$\cdot$yr.
Apart from the increased statistics also systematic uncertainties have
been improved: Thanks to the full volume calibration the error on the
fiducial mass has been reduced from 4.7\% to 1.8\%. Details of our
KamLAND analysis are described in appendix~A of
Ref.~\cite{Maltoni:2004ei}.  We use the data binned in equal bins in
$1/E$ to make optimal use of spectral information, we take into
account the (small) matter effect and carefully include various
systematics following Ref.~\cite{Huber:2004xh}. As previously, we
restrict the analysis to the prompt energy range above 2.6~MeV where
the contributions from geo-neutrinos~\cite{learned-nu08:07} as well as
backgrounds are small and the selection efficiency is roughly
constant~\cite{:2008ee}. In that energy range 1549 reactor neutrinos
events and a background of 63 events are expected without
oscillations, whereas the observed number of events is
985~\cite{kamland-taup07}.

The Sudbury Neutrino Observatory (SNO) has released the data of its
last phase, where the neutrons produced in the neutrino NC interaction
with deuterium are detected mainly by an array of $^3$He proportional
counters to measure the rate of neutral-current interactions in heavy
water and precisely determine the total active boron solar neutrino
flux, yielding the result
$5.54^{+0.33}_{-0.31}{\rm(stat)}^{+0.36}_{-0.34}{\rm(syst)} \times
10^6 \, {\rm cm}^{-2} {\rm s}^{-1}$~\cite{Aharmim:2008kc}.
The independent $^3$He neutral current detectors (NCD) provide a measurement
of the neutral current flux uncorrelated with the charged current rate
from solar $\nu_e$, different from the statistical CC/NC separation of
previous SNO phases. Since the total NC rate receives contributions
from the NCD as well as from the PMTs (as previously) a small
(anti-) correlation between CC and NC remains. Following
Ref.~\cite{Fogli:2008jx} we assume a correlation of $\rho = -0.15$.
In our SNO analysis we add the new data on the CC and NC fluxes to the
previous results~\cite{Aharmim:2005gt} assuming no correlation between
the NCD phase and the previous salt phase, see
Ref.~\cite{Maltoni:2003da} for further details. The main impact of the
new SNO data is due to the lower value for the observed CC/NC ratio,
$(\phi_\mathrm{CC}/\phi_\mathrm{NC})^\mathrm{NCD} =
0.301\pm0.033$~\cite{Aharmim:2008kc}, compared to the previous value
$(\phi_\mathrm{CC}/\phi_\mathrm{NC})^\mathrm{salt} =
0.34\pm0.038$~\cite{Aharmim:2005gt}.  Since for $^8$B neutrinos
$\phi_\mathrm{CC}/\phi_\mathrm{NC} \approx P_{ee} \approx
\sin^2\theta_{12}$, adding the new data point on this ratio with the
lower value leads to a stronger upper bound on $\sin^2\theta_{12}$.

 
We also include the direct measurement of the $^7$Be solar neutrino
signal rate performed by the Borexino
collaboration~\cite{Collaboration:2008mt}. They report an interaction
rate of the 0.862 MeV $^7$Be neutrinos of 49$\pm$3(stat)$\pm$4(syst)
counts/(day$\cdot$100~ton).
This measurement constitutes the first direct determination of the
survival probability for solar $\nu_e$ in the transition region
between matter-enhanced and vacuum-driven oscillations. The survival
probability of 0.862~MeV $^7$Be neutrinos is determined to be
$P_{ee}^\mathrm{^7Be,obs} = 0.56 \pm 0.1$. We find that with present
errors Borexino plays no significant role in the determination of
neutrino oscillation parameters. Apart from the fact that the
uncertainty on the survival probability is about a factor 3 larger
than e.g., the uncertainty on the SNO CC/NC ratio measurement, it turns
out that the observed value for $P_{ee}$ quoted above practically
coincides (within $0.1\sigma$) with the prediction at the best fit
point: $P_{ee}^\mathrm{^7Be,pred} = 0.55$.

The new data from SNO and Borexino are combined with the global data
on solar neutrinos~\cite{Cleveland:1998nv, sage, Altmann:2005ix,
Hosaka:2005um}, where we take into account the results of a recent
re-analysis of the Gallex data yielding a combined Gallex and GNO rate
of $67.6 \pm 4.0 \pm3.2$~SNU~\cite{gallex-nu08:07}.

\begin{figure}
 \centering
 \includegraphics[width=0.8\textwidth]{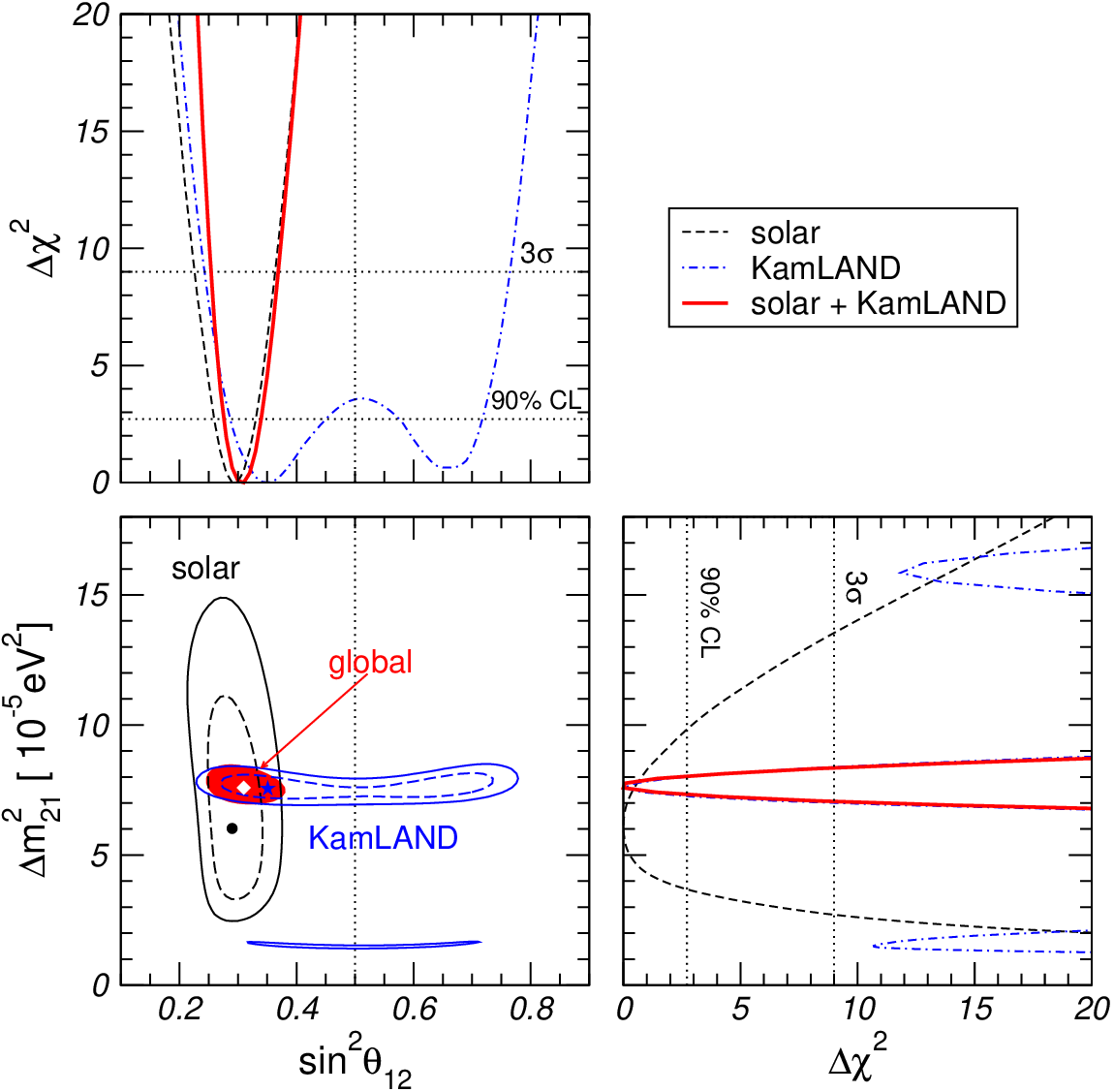}
 \caption{Determination of the leading ``solar'' oscillation
   parameters from the interplay of data from artificial and natural
   neutrino sources. We show $\chi^2$-profiles and allowed regions at
   90\% and 99.73\%~CL (2~dof) for solar and KamLAND, as well as the
   99.73\%~CL region for the combined analysis. The dot, star and
   diamond indicate the best fit points of solar data, KamLAND and
   global data, respectively. We minimise with respect to $\Dma$,
   $\theta_{23}$ and $\theta_{13}$, including always atmospheric,
   MINOS, K2K and CHOOZ data.}
  \label{fig:dominant-12}
\end{figure}

Fig.~\ref{fig:dominant-12} illustrates how the determination of the
leading solar oscillation parameters $\theta_{12}$ and $\Dmq_{21}$
emerges from the complementarity of solar and reactor neutrinos. From
the global three-flavour analysis we find ($1\sigma$ errors)
\begin{equation}\label{eq:solar}
\sin^2\theta_{12} = 0.304^{+0.022}_{-0.016} \,,\qquad
\Delta m^2_{21} = 7.65^{+0.23}_{-0.20} \times 10^{-5}\,\eVq \,.
\end{equation}
The numerical $\chi^2$ profiles shown in Fig.~\ref{fig:dominant-12}
have to very good accuracy the Gaussian shape $\chi^2 =
(x-x_\mathrm{best})^2/\sigma^2$, when the different $\sigma$ for upper
an lower branches are used as given in Eq.~(\ref{eq:solar}). Spectral
information from KamLAND data leads to an accurate determination of
$\Delta m^2_{21}$ with the remarkable precision of 8\% at $3\sigma$,
defined as $(x^\mathrm{upper} - x^\mathrm{lower})/(x^\mathrm{upper} +
x^\mathrm{lower})$.  We find that the main limitation for the $\Dms$
measurement comes from the uncertainty on the energy scale in KamLAND
of 1.5\%.  KamLAND data start also to contribute to the lower bound on
$\sin^2\theta_{12}$, whereas the upper bound is dominated by solar
data, most importantly by the CC/NC solar neutrino rate measured by
SNO. The SNO-NCD measurement reduces the $3\sigma$ upper bound on
$\sin^2\theta_{12}$ from 0.40~\cite{Maltoni:2004ei} to 0.37.

\bigskip 

Let us now move to the discussion of the status of the leading
atmospheric parameters $\theta_{23}$ and $\Dma$. The Main Injector
Neutrino Oscillation Search experiment (MINOS) has reported new
results on $\nu_\mu$ disappearance with a baseline of 735~km based on
a two-year exposure from the Fermilab NuMI beam. Their data, recorded
between May 2005 and July 2007 correspond to a total of $3.36 \times
10^{20}$ protons on target (POT)~\cite{Adamson:2008zt}, more than
doubling the POT with respect to MINOS run~I~\cite{Michael:2006rx},
and increasing the exposure used in the latest version of
Ref.~\cite{Maltoni:2004ei} by about 34\%.
The latest data confirm the energy dependent disappearance of
$\nu_\mu$, showing significantly less events than expected in the case
of no oscillations in the energy range $\lesssim 6$~GeV, whereas the
high energy part of the spectrum is consistent with the no oscillation
expectation. We include this result in our analysis by fitting the
event spectrum given in Fig.~2 of Ref.~\cite{Adamson:2008zt}. Current
MINOS data largely supersedes the pioneering K2K
measurement~\cite{Aliu:2004sq} which by now gives only a very minor
contribution to the $\Dma$ measurement.

We combine the long-baseline accelerator data with atmospheric
neutrino measurements from Super-Kamiokande~\cite{Ashie:2005ik}, using
the results of Ref.~\cite{Maltoni:2004ei}, see references therein for
details. In this analysis sub-leading effects of $\Dms$ in atmospheric
data are neglected, but effects of $\theta_{13}$ are included, in a
similar spirit as in Ref.~\cite{Hosaka:2006zd}.

\begin{figure}
  \centering
  \includegraphics[width=0.8\textwidth]{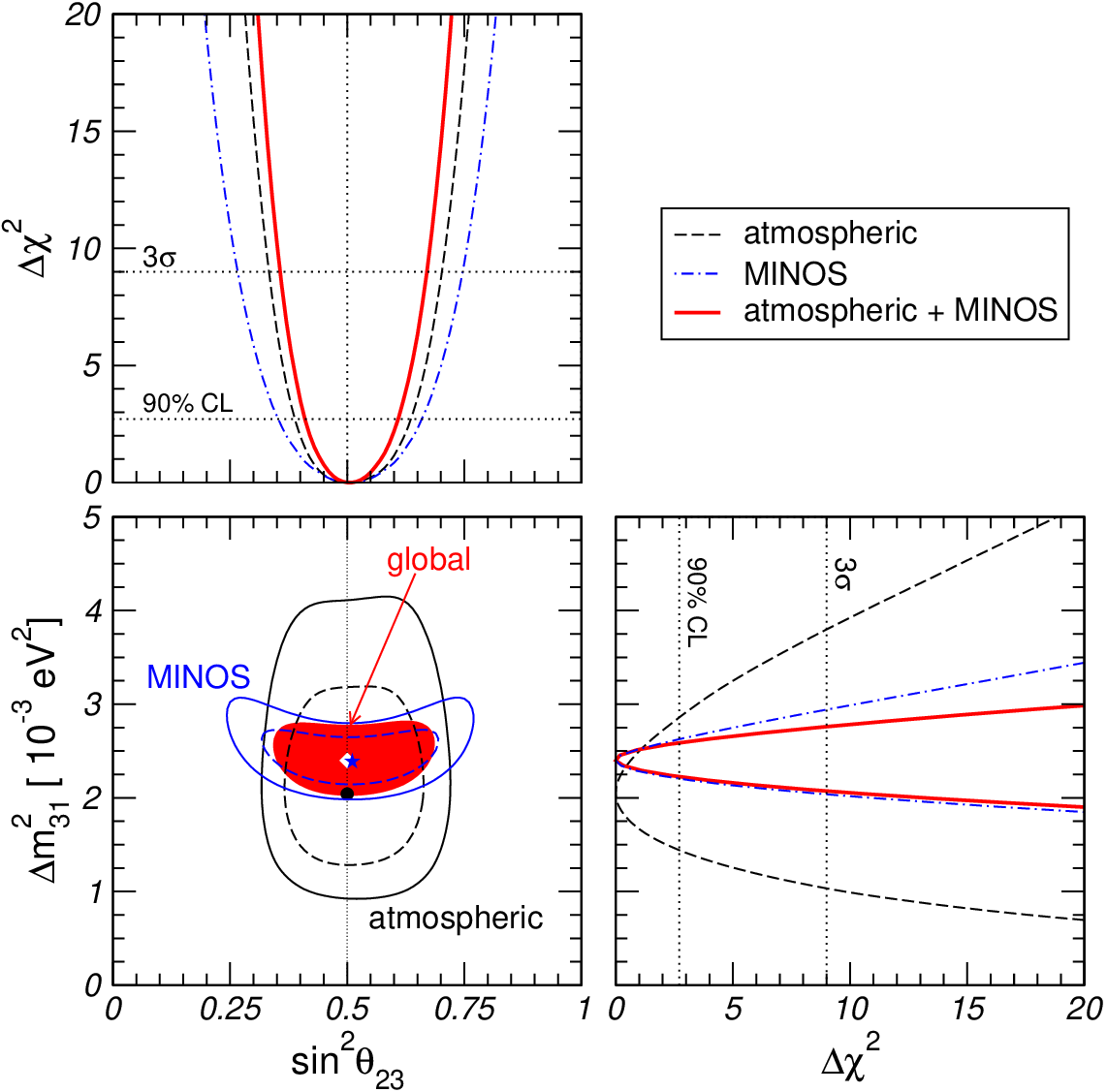}
  \caption{Determination of the leading ``atmospheric'' oscillation
    parameters from the interplay of data from artificial and natural
    neutrino sources. We show $\chi^2$-profiles and allowed regions at
    90\% and 99.73\%~CL (2~dof) for atmospheric and MINOS, as well as
    the 99.73\%~CL region for the combined analysis (including also
    K2K). The dot, star and diamond indicate the best fit points of
    atmospheric data, MINOS and global data, respectively. We minimise
    with respect to $\Dms$, $\theta_{12}$ and $\theta_{13}$, including
    always solar, KamLAND, and CHOOZ data.}
  \label{fig:dominant-23}
\end{figure}

Fig.~\ref{fig:dominant-23} illustrates how the determination of the
leading atmospheric oscillation parameters $\theta_{23}$ and
$|\Dmq_{31}|$ emerges from the complementarity of atmospheric and
accelerator neutrino data. We find the following best fit points and
$1\sigma$ errors:
\begin{equation}
\sin^2\theta_{23} = 0.50^{+0.07}_{-0.06} \,,\qquad
|\Delta m^2_{31}| = 2.40^{+0.12}_{-0.11} \times 10^{-3}\,\eVq \,.
\end{equation}
The determination of $|\Dma|$ is dominated by spectral data from the
MINOS long-baseline $\nu_\mu$ disappearance experiment, where the sign
of $\Dmq_{31}$ (i.e., the neutrino mass hierarchy) is undetermined by
present data. The measurement of the mixing angle $\theta_{23}$ is
still largely dominated by atmospheric neutrino data from
Super-Kamiokande with a best fit point at maximal mixing. Small
deviations from maximal mixing due to sub-leading three-flavour
effects have been found in
Refs.~\cite{Fogli:2005cq,GonzalezGarcia:2007ib}, see, however, also
Ref.~\cite{kajita} for a preliminary analysis of Super-Kamiokande. A
comparison of these subtle effects can be found in
Ref.~\cite{snow}. At present deviations from maximality are not
statistically significant.

\section{Status of $\theta_{13}$}
\label{sec:th13}

The third mixing angle $\theta_{13}$ would characterize the magnitude
of CP violation in neutrino oscillations.  Together with the
determination of the neutrino mass spectrum hierarchy (i.e., the sign
of $\Dma$) it constitutes a major open challenge for any future
investigation of neutrino
oscillations~\cite{Bandyopadhyay:2007kx,Nunokawa:2007qh}.

\begin{figure}
  \centering
  \includegraphics[height=7cm]{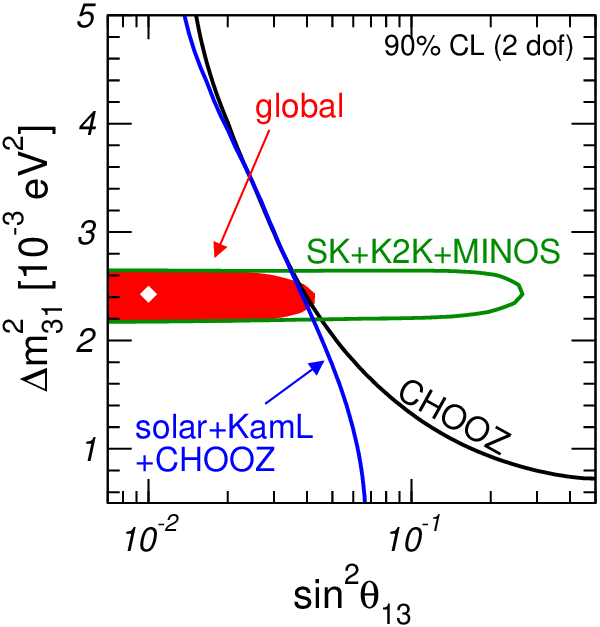}
  \quad
  \includegraphics[height=7cm]{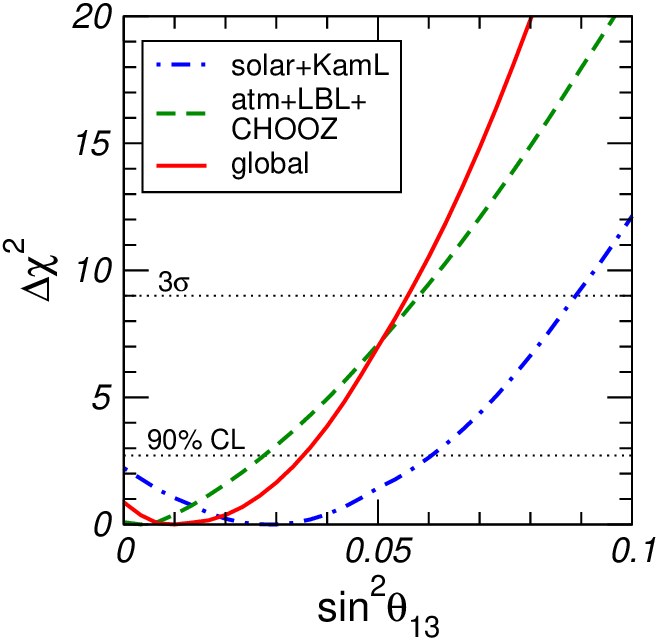}
  \caption{Constraints on $\sin^2\theta_{13}$ from different parts of
  the global data.}
  \label{fig:th13}
\end{figure}

Fig.~\ref{fig:th13} summarizes the information on $\theta_{13}$ from
present data. Similar to the case of the leading oscillation
parameters, also the bound on $\theta_{13}$ emerges from an interplay
of different data sets, as illustrated in the left panel of
Fig.~\ref{fig:th13}. An important contribution to the bound comes, of
course, from the CHOOZ reactor experiment~\cite{Apollonio:2002gd}
combined with the determination of $|\Dmq_{31}|$ from atmospheric and
long-baseline experiments. Due to a complementarity of low and high
energy solar neutrino data, as well as solar and KamLAND data, one
finds that also solar+KamLAND provide a non-trivial constraint on
$\theta_{13}$, see e.g.,
Refs.~\cite{Maltoni:2003da,Maltoni:2004ei}~\cite{Goswami:2004cn}.  We
obtain at 90\%~CL ($3\sigma$) the following limits~\footnote{Note that
the bounds given in Eq.~(\ref{eq:th13}) are obtained for 1~dof,
whereas in Fig.~\ref{fig:th13} (left) the 90\%~CL regions for 2~dof
are shown.}:
\begin{equation}\label{eq:th13}
  \sin^2\theta_{13} \le \left\lbrace \begin{array}{l@{\qquad}l}
      0.060~(0.089) & \text{(solar+KamLAND)} \\
      0.027~(0.058) & \text{(CHOOZ+atm+K2K+MINOS)} \\
      0.035~(0.056) & \text{(global data)}
    \end{array} \right.
\end{equation}
In the global analysis we find a slight weakening of the upper bound
on $\sin^2\theta_{13}$ due to the new data from 0.04 (see
Ref.~\cite{snow} or v5 of \cite{Maltoni:2004ei}) to 0.056 at
$3\sigma$. The reason for this is two-fold. First, the shift of the
allowed range for $|\Delta m^2_{31}|$ to lower values due to the new
MINOS data implies a slightly weaker constraint on $\sin^2\theta_{13}$
(cf.\ Fig.~\ref{fig:th13}, left), and second, the combination of solar
and new KamLAND data prefers a slightly non-zero value of
$\sin^2\theta_{13}$ which, though not statistically significant, also
results in a weaker constraint in the global fit (cf.\
Fig.~\ref{fig:th13}, right).

\begin{figure}
  \centering
  \includegraphics[height=7cm]{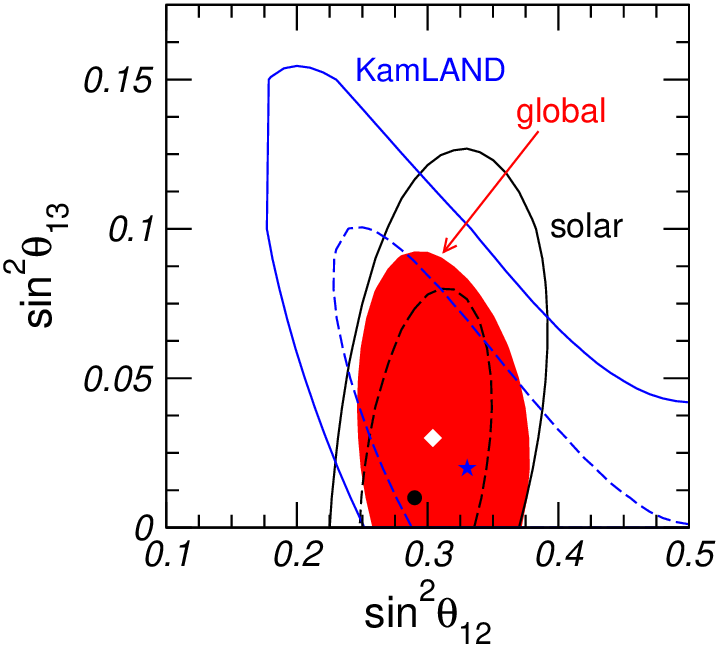}
  \quad
  \includegraphics[height=7cm]{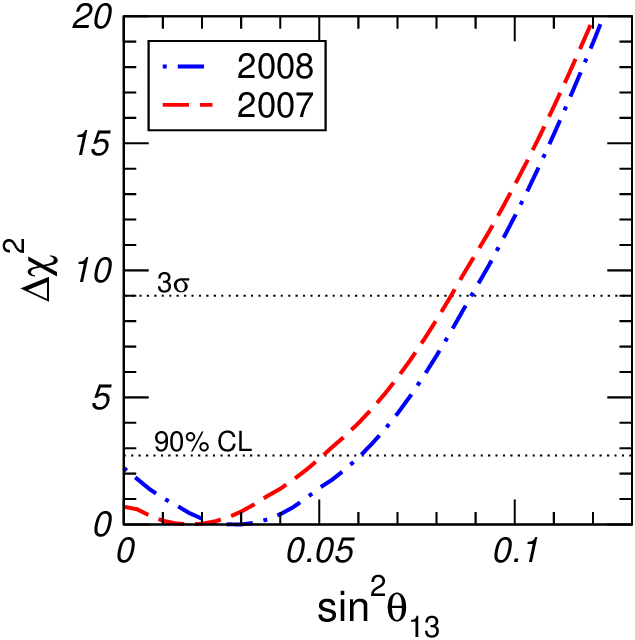}
  \caption{Left: Allowed regions in the $(\theta_{12}-\theta_{13})$
   plane at 90\% and 99.73\%~CL (2~dof) for solar and KamLAND, as well
   as the 99.73\%~CL region for the combined analysis. $\Dms$ is fixed
   at its best fit point. The dot, star, and diamond indicate the
   best fit points of solar, KamLAND, and combined data,
   respectively. Right: $\chi^2$ profile from solar and KamLAND data
   with and without the 2008 SNO-NCD data.}
  \label{fig:th13-sol-kl}
\end{figure}

As has been noted in Ref.~\cite{Fogli:2008jx} the slight downward
shift of the SNO CC/NC ratio due to the SNO-NCD data leads to a ``hint''
for a non-zero value of $\theta_{13}$. From the combination of solar
and KamLAND data we find a best fit value of $\sin^2\theta_{13} =
0.03$ with $\Delta\chi^2 = 2.2$ for $\theta_{13} = 0$ which
corresponds to a $1.5\sigma$ effect (86\%~CL). 
We illustrate the interplay of solar and KamLAND data in the left
panel of Fig.~\ref{fig:th13-sol-kl}. The survival probability in
KamLAND is given by 
\begin{equation}
P_{ee}^\mathrm{KamL} \approx
\cos^4\theta_{13} \left(1-\sin^22\theta_{12}\sin^2\frac{\Dms\ L}{4E}\right) \,,
\end{equation}
leading to an anti-correlation of $\sin^2\theta_{13}$ and
$\sin^2\theta_{12}$~\cite{Maltoni:2004ei}, see
also~\cite{Goswami:2004cn,Balantekin:2008zm}. In contrast, for 
solar neutrinos one has
\begin{equation}\label{eq:sun}
P_{ee}^\mathrm{solar} \approx
\left\{\begin{array}{l@{\qquad}l}
\cos^4\theta_{13} \left(1- \frac{1}{2}\sin^22\theta_{12}\right) & 
\mbox{low energies} \\
\cos^4\theta_{13} \, \sin^2\theta_{12} & 
\mbox{high energies} 
\end{array}\right. \,.
\end{equation}
Eq.~(\ref{eq:sun}) shows a similar anti-correlation as in KamLAND for the
vacuum oscillations of low energy solar neutrinos. For the high energy
part of the spectrum, which undergoes the adiabatic MSW conversion
inside the sun and which is subject to the SNO CC/NC measurement, a
positive correlation of $\sin^2\theta_{13}$ and $\sin^2\theta_{12}$
emerges. As visible from Fig.~\ref{fig:th13-sol-kl}~(left) and as
discussed already in Refs.~\cite{Maltoni:2004ei,Goswami:2004cn}, this
complementarity leads to a non-trivial constraint on $\theta_{13}$ and
it allows to understand the hint for a non-zero value of
$\theta_{13}$, which helps to reconcile the slightly different best
fit points for $\theta_{12}$ for solar and KamLAND
separately~\cite{Balantekin:2008zm,Fogli:2008jx}.
This trend was visible already in pre-SNO-NCD data, though with a
significance of only $0.8\sigma$, see
Fig.~\ref{fig:th13-sol-kl}~(right) showing the present result together
with our previous one from v6 of~\cite{Maltoni:2004ei}.

Let us briefly comment on a possible additional hint for a non-zero
$\theta_{13}$ from atmospheric neutrino data~\cite{Fogli:2005cq,
Escamilla:2008vq}; Refs.~\cite{Fogli:2005cq, Fogli:2008jx} find from
atmospheric+long-baseline+CHOOZ data a $0.9\sigma$ hint for a non-zero
value: $\sin^2\theta_{13} = 0.012 \pm 0.013$. In our atmospheric
neutrino analysis (neglecting $\Dms$ effects) combined with CHOOZ data
the best fit occurs for $\theta_{13} = 0$ (cf.\ Fig.~\ref{fig:th13},
right), in agreement with Ref.~\cite{Hosaka:2006zd}. Also, in the
atmospheric neutrino analysis from Ref.~\cite{GonzalezGarcia:2007ib}
(which does include $\Dms$ effects, as Refs.~\cite{Fogli:2005cq,
Fogli:2008jx}) the preference for a non-zero $\theta_{13}$ is much
weaker than the one from~\cite{Fogli:2005cq}, with a $\Delta\chi^2
\lesssim 0.2$. 
In our global analysis the hint from solar+KamLAND gets diluted by the
constraint coming from atmospheric+CHOOZ data, and we find the global
$\chi^2$ minimum at $\sin^2\theta_{13} = 0.01$, but with $\theta_{13}
= 0$ allowed at $0.9\sigma$ ($\Delta\chi^2 = 0.87$). Hence, we
conclude that at present there is no significant hint for a non-zero
$\theta_{13}$. As already stated, the origin of slightly different
conclusions of other studies is related with including or neglecting
the effect of solar terms in the atmospheric neutrino oscillation
analysis, and translates also into a possibly nonmaximal best fit value for
$\theta_{23}$. Note, however, that all analyses agree within
$\Delta\chi^2$ values of order 1 and therefore there is no significant
disagreement. A critical discussion of the impact of sub-leading
effects in atmospheric data on $\theta_{13}$ and $\theta_{23}$ as well
as a comparison of the results of different groups can be found in
Ref.~\cite{snow}.

Before summarizing let us update also the determination of the 
ratio of the two mass-squared differences,
\begin{equation}
  \alpha \equiv \frac{\Delta m^2_{21}}{|\Delta m^2_{31}|} = 
0.032\,, \quad 0.027 \le \alpha \le 0.038 \quad (3\sigma) \,,
\end{equation}
which is relevant for the description of CP violation in neutrino
oscillations in long-baseline experiments.

\section{Summary}
\label{sec:summary}

In this work we have provided an update on the status of three-flavour
neutrino oscillations, taking into account the latest available world
neutrino oscillation data presented at the {\em Neutrino 2008\/}
Conference. Our results are summarized in
Figures~\ref{fig:dominant-12}, \ref{fig:dominant-23} and
\ref{fig:th13}.  Table~\ref{tab:summary} provides best fit points,
$1\sigma$ errors, and the allowed intervals at 2 and 3$\sigma$ for the
three-flavour oscillation parameters.

\begin{table}[ht]\centering
    \catcode`?=\active \def?{\hphantom{0}}
    
\begin{tabular}{|@{\quad}>{\rule[-2mm]{0pt}{6mm}}l@{\quad}|@{\quad}c@{\quad}|@{\quad}c@{\quad}|@{\quad}c@{\quad}|}
        \hline
        parameter & best fit & 2$\sigma$ & 3$\sigma$ 
        \\
        \hline
        $\Delta m^2_{21}\: [10^{-5}\eVq]$
        & $7.65^{+0.23}_{-0.20}$  & 7.25--8.11 & 7.05--8.34 \\[2mm]
        $|\Delta m^2_{31}|\: [10^{-3}\eVq]$
        & $2.40^{+0.12}_{-0.11}$  & 2.18--2.64 & 2.07--2.75 \\[2mm]
        $\sin^2\theta_{12}$
        & $0.304^{+0.022}_{-0.016}$ & 0.27--0.35 & 0.25--0.37\\[2mm]
        $\sin^2\theta_{23}$
        & $0.50^{+0.07}_{-0.06}$ & 0.39--0.63 & 0.36--0.67\\[2mm]
        $\sin^2\theta_{13}$
        & $0.01^{+0.016}_{-0.011}$  & $\leq$ 0.040 & $\leq$ 0.056 \\
        \hline
\end{tabular}
\caption{ \label{tab:summary} Best-fit values with 1$\sigma$ errors,
  and 2$\sigma$ and 3$\sigma$ intervals (1 \dof) for the
  three--flavour neutrino oscillation parameters from global data
  including solar, atmospheric, reactor (KamLAND and CHOOZ) and
  accelerator (K2K and MINOS) experiments.}
\end{table}

\paragraph{Acknowledgments.} This work was supported by MEC grant
FPA2005-01269, by EC Contracts RTN network MRTN-CT-2004-503369 and
ILIAS/N6 RII3-CT-2004-506222.  We thank Michele Maltoni for
collaboration on global fits to neutrino oscillation data.

\appendix\section{Updated analysis as of February 2010}

\subsection{Updates in the solar neutrino analysis}
\label{sec:solar-update}

\underline{SSM}: 
We consider the recently updated standard solar model from
\cite{Serenelli:2009yc}. Among the different models presented in that
reference, we use the low metallicity model, labelled as AGSS09, that
incorporates the most recent determination of solar
abundances~\cite{Asplund:2009fu} as our standard choice. The solar
abundances in that model are a bit higher than previous determinations by
the same group, alleviating the disagreement with helioseismic data.  From
the point of view of solar neutrinos, the most important changes with
respect to the previous SSM used in our analysis (BS05(OP), with high
metallicities~\cite{Bahcall:2004pz}) is the 15\% and 5\% reduction in the
Boron and Beryllium fluxes respectively. This is due to the reduced central
temperature in the new model with respect to the previous one. Given the
condition of fixed solar luminosity, this reduction is compensated by a
slight increase in the pp and pep neutrino fluxes. We discuss also the
impact of a new SSM with high metallicity, the GS98 model (presented in
\cite{Serenelli:2009yc} as well), see also the recent discussion
in~\cite{GonzalezGarcia:2010er}.

\noindent
\underline{SAGE}:
In our present analysis we have updated the capture rate of solar neutrinos
measured in SAGE: $65.4{^{+3.1}_{-3.0}}{^{+2.6}_{-2.8}}$~SNU
\cite{Abdurashitov:2009tn}, compared to the previous value
$66.9{^{+3.9}_{-3.8}}{^{+3.6}_{-3.2}}$~SNU \cite{sage-old}. Note that the
recently published reanalysis of GALLEX data~\cite{Kaether:2010ag} has been
reported already at Neutrino 2008~\cite{gallex-nu08:07} and was therefore
included in the original version of this paper.

\noindent
\underline{SNO}:
In our update we include also the results from the recent joint re-analysis of
data from the Phase I and Phase II (the pure D$_2$O and salt phases) of the
Sudbury Neutrino Observatory (SNO)~\cite{Collaboration:2009gd}. In this
analysis, an effective electron kinetic energy threshold of 3.5~MeV has been
used (Low Energy Threshold Analysis, LETA), and the total flux of $^8$B
neutrinos has been determined to be 
\begin{equation}
\phi_{NC} = 5.140 ^{+0.160}_{-0.158} {\rm (stat)} ^{+0.132}_{-0.117}
{\rm (syst)}  \times 10^6 \, \rm cm^{-2} s^{-1} \,.
\end{equation}
Comparing this number with the result obtained in SNO Phase III (the NC
detector phase, NCD):
$5.54^{+0.33}_{-0.31}{\rm(stat)}^{+0.36}_{-0.34}{\rm(syst)} \times 10^6 \,
{\rm cm}^{-2} {\rm s}^{-1}$ \cite{Aharmim:2008kc}, one can see that the
determination of the total neutrino flux has been improved by about a factor
2.  These improvements have been possible thanks mainly to the increased
statistics, in particular the NC event sample in the LETA is increased by
about 70\%, since the previously used higher energy thresholds of 5~MeV in
phase~I and 5.5~MeV in phase~II have cut away a significant portion of the
NC events. Furthermore, energy resolution, backgrounds suppression, and
systematic uncertainties have been improved.
We include the LETA SNO data by fitting the predicted energy-dependent
neutrino survival probability and day-night asymmetry in terms of the
polynomials given by the SNO collaboration, see Tabs.~XXVI and XXVII
in~\cite{Collaboration:2009gd}.
We have checked that our results agree with the analysis including all solar
neutrino experiments made by SNO. 
Note that we have adopted in our present analysis of the SNO-NCD phase
data the detailed correlations between the CC, NC and ES neutrino
fluxes recently given by the SNO collaboration~\cite{sno-ncd-howto},
thereby improving our previous treatment presented in Sec.~2.

\begin{figure}
  \centering
  \includegraphics[width=\textwidth]{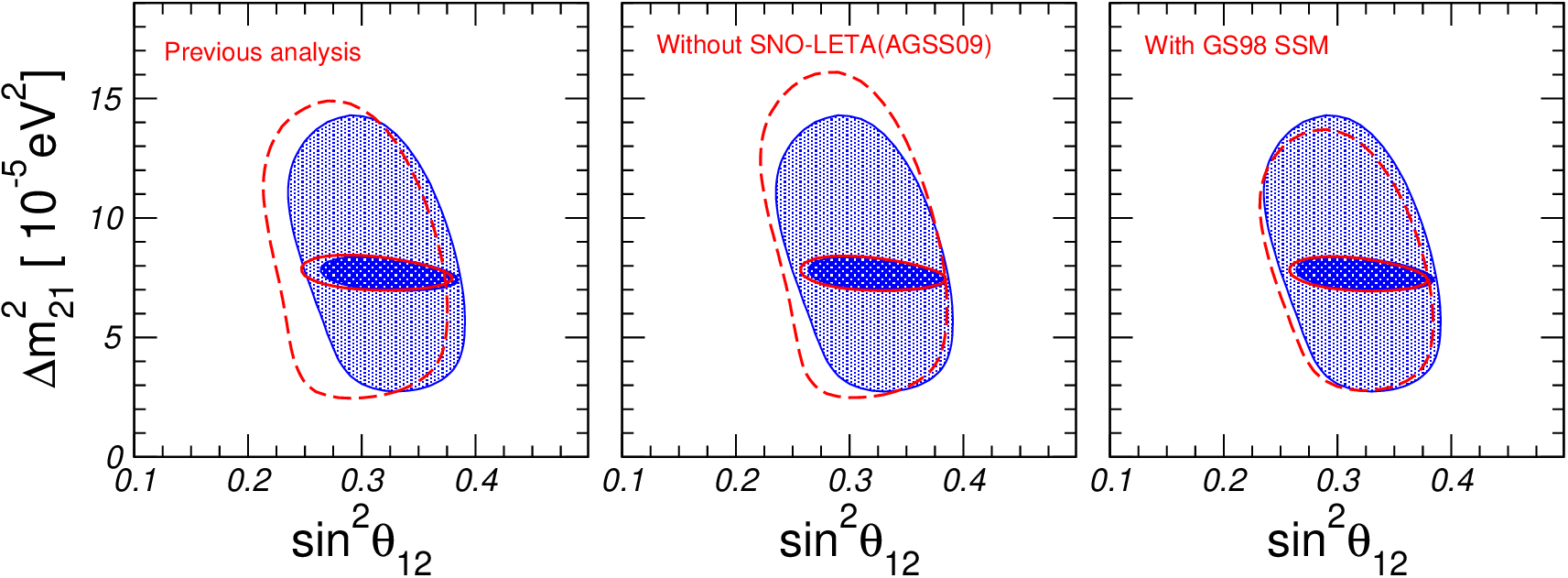}
  \caption{Impact of the changes in the solar neutrino analysis. In all
  panels the blue (shaded) regions corresponds to the $3\sigma$ regions from
  solar and solar+KamLAND updated analysis. The regions delimited by the red
  contour curves correspond to our previous analysis (left), an analysis
  using the previous high-threshold SNO phase I and II analysis but the same
  solar model (middle), and an analysis using the high metallicity GS98
  instead of our standard low metallicity AGSS09 solar model (right).
  \label{fig:solar-changes}}
\end{figure}

Fig.~\ref{fig:solar-changes} shows the impact of the updates in the solar
analysis on the determination of the solar parameters. The left panel
compares solar and solar+KamLAND allowed regions for the previous and
updated analyses, the middle panel illustrates the impact of the SNO LETA
analysis, and the right panel shows the effect of changing between the low
(AGSS09) and high (GS98) metallicity solar models. We observe that the main
changes come from the SNO LETA analysis, whereas the impact of solar
metallicity is small. In general changes are rather small, once KamLAND data
is added to the solar data, with a small tightening of the lower bound on
$\sin^2\theta_{12}$. The best fit point value for the solar mixing angle has
been shifted to a slightly higher value mainly due to the lower values
reported for the total boron neutrino flux, either from the SNO LETA
measurements as well as from the updated SSM with low metallicities. The
allowed range for $\Delta m^2_{21}$ of solar only data has been somewhat
reduced, however this effect gets completely diluted after combining with
KamLAND data.
The updated best fit values and allowed ranges for $\sin^2\theta_{12}$ and
$\Dms$ can be found in Tab.~\ref{tab:summary2010}.

\begin{figure}
  \centering
 \includegraphics[height=0.36\textwidth]{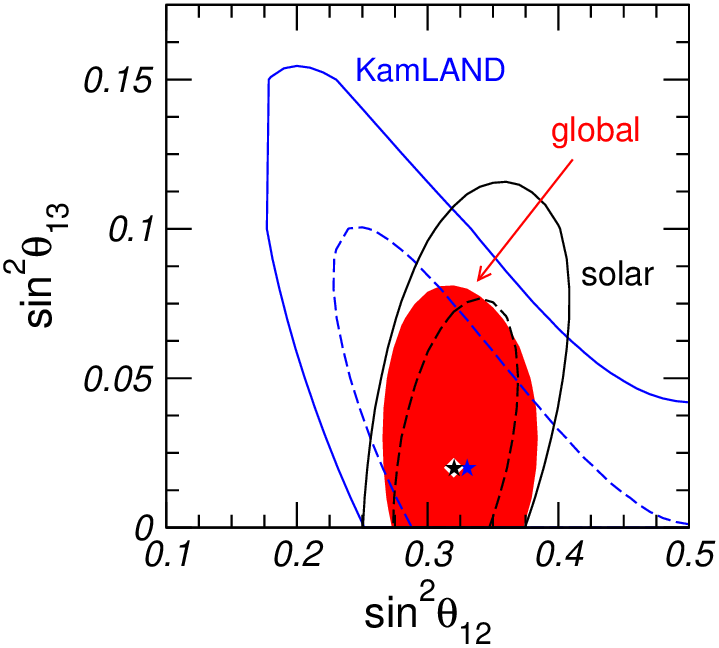}
 \quad
  \includegraphics[height=0.36\textwidth]{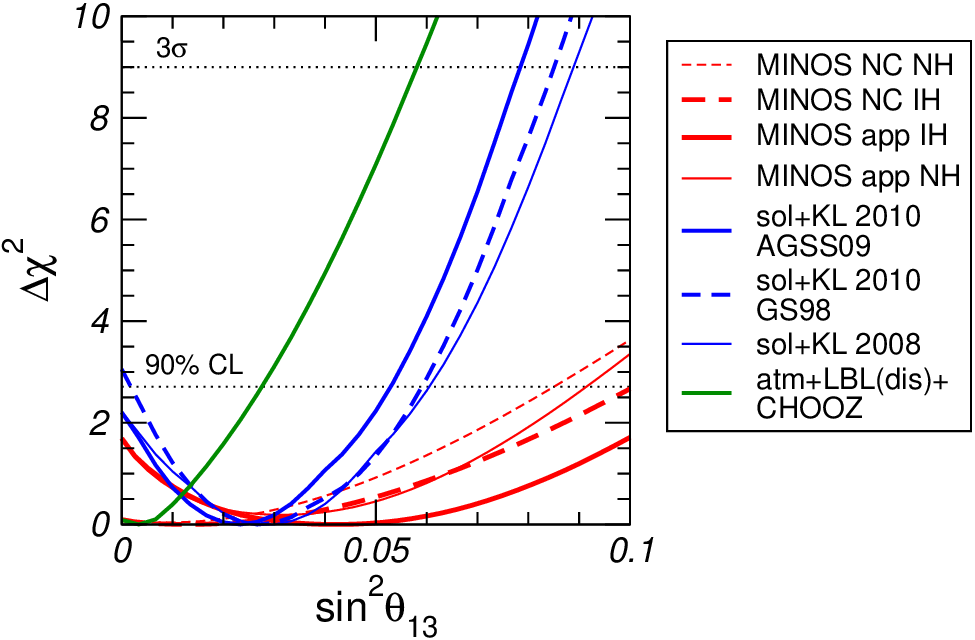}
  \caption{Left (update of Fig.~\ref{fig:th13-sol-kl}, left): Allowed
   regions in the $(\theta_{12}-\theta_{13})$ plane at 90\% and 99.73\%~CL
   (2~dof) for solar and KamLAND, as well as the 99.73\%~CL region for the
   combined analysis. Right: $\Delta\chi^2$ as a function of
   $\sin^2\theta_{13}$. The blue curves illustrate the impact of the updates
   in the solar neutrino analysis on the bound from the global solar+KamLAND
   data. The red curves show the constraint coming from the MINOS $\nu_e$
   appearance (red solid) and NC (red dashed) data, where we show the
   $\Delta\chi^2$ assuming NH (thin) and IH (thick), both with respect to
   the common minimum, which occurs for IH. The green solid curve corresponds
   the bound from CHOOZ + atmospheric + K2K + MINOS~(disappearance)
   data.\label{fig:th13-changes}}
\end{figure}

The impact of the updated solar analysis on $\theta_{13}$ is illustrated in 
Fig.~\ref{fig:th13-changes}. Our analysis of solar + KamLAND data gives $\sin^2\theta_{13} =
0.022^{+0.018}_{-0.015}$ in excellent agreement with the value obtained by
the SNO Collaboration~\cite{Collaboration:2009gd}, $\sin^2\theta_{13} =
0.0200^{+0.0209}_{-0.0163}$.
Hence, we obtain a lower best fit value with respect to the one we
obtained in our previous analysis ($\sin^2\theta_{13}$ = 0.03). This
is due to the fact that now solar data prefer a somewhat higher value
of $\theta_{12}$ (as KamLAND does), and therefore, a smaller value of
$\theta_{13}$ is required to reconcile solar and KamLAND data, as can
be seen by comparing left panels of Figs.~\ref{fig:th13-sol-kl} and
\ref{fig:th13-changes}.
The fact that now solar data prefer a larger value for
$\sin^2\theta_{12}$ results in a stronger bound on $\theta_{13}$ from
the combination of solar + KamLAND data. The allowed solar region in
the panel ($\sin^2\theta_{12}$, $\sin^2\theta_{13}$) is more shifted
to the right (because of the higher $\theta_{12}$ preferred by the new
smaller boron neutrino flux), where the allowed KamLAND region is
narrower. At $\theta_{13}=0$ we find $\Delta \chi^2 = 2.2$, same value
as before. 

As stated above, the small improvement in the $\theta_{13}$ bound is related
to the solar model used. For models with higher solar metallicities like
GS98, a slightly weaker bound is obtained~\cite{GonzalezGarcia:2010er}, see
Fig.~\ref{fig:th13-changes} (right). In that case we obtain a slightly
larger best fit point, $\sin^2\theta_{13} = 0.027^{+0.019}_{-0.015}$ and 
$\Delta \chi^2 = 3.05$ at $\theta_{13} = 0$.

\subsection{MINOS $\nu_e$ appearance data}

In Ref.~\cite{Collaboration:2009yc} a search for $\nu_\mu\to\nu_e$
transitions by the MINOS experiment has been presented, based on a $3.14
\times 10^{20}$ protons-on-target exposure in the Fermilab NuMI beam.  35
events have been observed in the far detector with a background of $27 \pm
5{\rm (stat)}\pm 2 {\rm (syst)}$ events predicted by the measurements in the
near detector. This corresponds to an excess of about $1.5\sigma$ which can
be interpreted as a weak hint for $\nu_e$ appearance due to a non-zero
$\theta_{13}$. We fit the MINOS $\nu_e$ spectrum by using the GLoBES
simulation software~\cite{Huber:2007ji}, where we calibrate our predicted
spectrum by using the information given in \cite{Boehm:2009zz}. A full
three-flavour fit is performed taking into account a 7.3\% uncertainty on
the background normalization (Tab.~I of ~\cite{Collaboration:2009yc}), and a
5\% uncertainty on the matter density along the neutrino path.

In the MINOS detector, being optimized for muons, it is rather
difficult to identify $\nu_e$~CC events, since they lead to an
electromagnetic shower.  NC and $\nu_\mu$~CC events often have a
similar signature, and hence lead to a background for the $\nu_e$
appearance search. Indeed, in Ref.~\cite{Adamson:2008jh} an analysis
of ``NC events'' has been performed, where ``NC events'' in fact
include also $\nu_e$~CC events due to the similar event
topology. Therefore, a possible $\nu_\mu \to\nu_e$ oscillation signal
would contribute to the ``NC event'' sample of~\cite{Adamson:2008jh}
and these data can be used to constrain $\theta_{13}$. We have
performed a fit to the observed spectrum, again using the GLoBES
software, by summing the NC events induced from the total neutrino
flux with the $\nu_e$~CC appearance signal due to oscillations. We
include a 4\% error on the predicted NC spectrum and a 3\% error on
the $\nu_\mu$ CC induced background (Tab.~II
of~\cite{Adamson:2008jh}).

In Fig.~\ref{fig:th13-changes} (right) we show the constraint on
$\sin^2\theta_{13}$ from these MINOS data. The $\chi^2$ has been
marginalized with respect to all parameters except $\theta_{13}$, where for
the solar and atmospheric parameters we imposed Gaussian errors taken from
Tab.~\ref{tab:summary2010}, without including any other information on
$\theta_{13}$ except from MINOS. We show the $\Delta\chi^2$ profiles for
$\nu_e$ appearance data and NC data, for a fixed neutrino mass hierarchy.
The best fit point is always obtained for the inverted hierarchy (IH,
$\Delta m^2_{31} < 0$), and in that case in general the constraint on
$\sin^2\theta_{13}$ is weaker, since for IH the matter effect tends to
suppress the $\nu_e$ appearance probability. The $\Delta\chi^2$ for normal
hierarchy (NH, $\Delta m^2_{31} > 0$) is given with respect to the best fit
for IH. In the global analysis we also marginalize over the two hierarchies,
and hence, the actual information from MINOS comes from the IH.

We see from the figure that MINOS $\nu_e$ appearance data shows a slight
preference for a non-zero value of $\theta_{13}$, with a best fit point of 
$\sin^2\theta_{13} = 0.032(0.043)$ for NH (IH) with $\Delta\chi^2 = 1.8$ at
$\sin^2\theta_{13} = 0$. In contrast, no indication for a non-zero
$\theta_{13}$ comes from the NC data. Furthermore, one observes that NC
gives a slightly more constraining upper bound on $\sin^2\theta_{13}$ than
$\nu_e$ appearance, while both are significantly weaker than the bound from
$\nu_\mu$ disappearance data + CHOOZ or solar+KamLAND. Let us mention that
the result for the NC analysis strongly depends on the value assumed for the
systematic uncertainty, whereas the $\nu_e$ appearance result is more robust
with respect to systematics, being dominated by statistics.

In the global analysis we do not combine the $\chi^2$'s from MINOS $\nu_e$
and NC data, since presumably the data are not independent and adding them
would imply a double counting of the same data. Therefore, we adopt the
conservative approach and use only $\nu_e$ appearance data without the
information from NC data in the global analysis. We have checked, however,
that adding both MINOS data sets leads to practically the same result in the
global fit, both for the ``hint'' for $\theta_{13} > 0$ as well as the
global bound, the latter being dominated by other data sets.

\subsection{Updated global three-flavour fit}

\begin{figure}
  \centering
  \includegraphics[height=7.5cm]{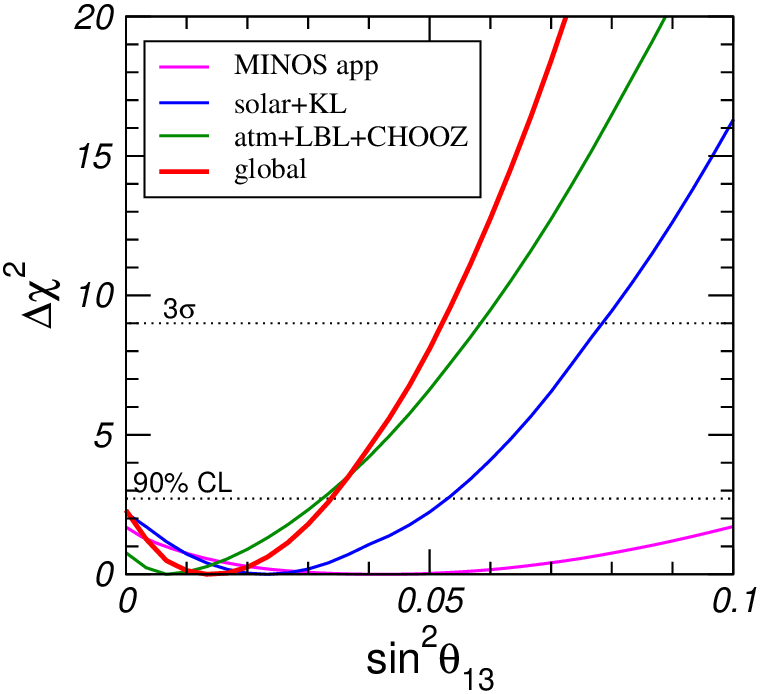}
  \caption{The constraint on $\sin^2\theta_{13}$ from MINOS $\nu_e$
  appearance data, solar + KamLAND data, atmospheric + CHOOZ + K2K +
  MINOS (disappearance as well as appearance), and the combined global
  data. \label{fig:th13-2010}}
\end{figure}

The present situation on the mixing angle $\theta_{13}$ is summarized
in Fig.~\ref{fig:th13-2010}. We obtain the following bounds at 90\%
($3\sigma$)~CL:
\begin{equation}\label{eq:th13-2010}
  \sin^2\theta_{13} \le \left\lbrace \begin{array}{l@{\qquad}l}
      0.053~(0.078) & \text{(solar+KamLAND)} \\
      0.033~(0.061) & \text{(CHOOZ+atm+K2K+MINOS)} \\
      0.034~(0.053) & \text{(global data)}
    \end{array} \right.
\end{equation}
We note a slight tightening of the bounds from solar+KamLAND as well
as the global bound, due to the update in the solar analysis, see
\ref{sec:solar-update}, whereas the bound from
CHOOZ+atm+K2K+MINOS gets slightly weaker, due to MINOS appearance
data.
In the global analysis we obtain the following best fit
value and 1$\sigma$ range:
\begin{equation}\label{eq:th13-1s-2010}
\sin^2\theta_{13} = 0.013^{+0.013}_{-0.009} 
\end{equation}
This corresponds to a $1.5\sigma$ hint for $\theta_{13} > 0$
($\Delta \chi^2 = 2.3$ at $\theta_{13}=0$).
As discussed in sec.~\ref{sec:th13} above, in our previous analysis
the 1.5$\sigma$ hint for $\theta_{13} > 0$ from solar+KamLAND data was
diluted after the combination with atmospheric, long-baseline and
CHOOZ data, resulting in a combined effect of 0.9$\sigma$. Now, thanks
to the new MINOS appearance data, we find that the atmospheric +
long-baseline + CHOOZ analysis already gives a nonzero best fit value
of $\theta_{13}$ (see Fig.~\ref{fig:th13-2010}), leading to the above
global result, eq.~\ref{eq:th13-1s-2010}.

Finally, let us comment on the possible hint for a non-zero $\theta_{13}$
from atmospheric data~\cite{Fogli:2005cq, Fogli:2008jx}, as discussed in
sec.~\ref{sec:th13}. The possible origin of such a hint has been
investigated in Ref.~\cite{Maltoni:2008ka} and recently in
\cite{GonzalezGarcia:2010er}, see also~\cite{Fogli:2009ce}.
From these results one may conclude that the
statistical relevance of the claimed hint for non-zero $\theta_{13}$ from
atmospheric data depends strongly on the details of the rate calculations
and of the $\chi^2$ analysis. Furthermore, the origin of that effect might
be traced back to a small excess (at the $1\sigma$ level) in the multi-GeV
$e$-like data sample in SK-I, which however, is no longer present in the
combined SK-I and SK-II, as well as SK-I+II+III data.

\begin{table}[ht]\centering
    \catcode`?=\active \def?{\hphantom{0}}    
    \begin{tabular}{|@{\quad}>{\rule[-2mm]{0pt}{6mm}}l@{\quad}|@{\quad}c@{\quad}|@{\quad}c@{\quad}|@{\quad}c@{\quad}|}
        \hline
        parameter & best fit & 2$\sigma$ & 3$\sigma$ 
        \\
        \hline
        $\Delta m^2_{21}\: [10^{-5}\eVq]$
        & $7.59^{+0.23}_{-0.18}$  & 7.22--8.03 & 7.03--8.27 \\[2mm] 
        $|\Delta m^2_{31}|\: [10^{-3}\eVq]$
        & $2.40^{+0.12}_{-0.11}$  & 2.18--2.64 & 2.07--2.75 \\[2mm] 
        $\sin^2\theta_{12}$
        & $0.318^{+0.019}_{-0.016}$ & 0.29--0.36 & 0.27--0.38\\[2mm]  
        $\sin^2\theta_{23}$
        & $0.50^{+0.07}_{-0.06}$ & 0.39--0.63 & 0.36--0.67\\[2mm] 
        $\sin^2\theta_{13}$
        & $0.013^{+0.013}_{-0.009}$  & $\leq$ 0.039 & $\leq$ 0.053 \\ 
        \hline
\end{tabular}
\caption{ \label{tab:summary2010} Current update of
  Tab.~\ref{tab:summary}: Best-fit values with 1$\sigma$ errors, and
  2$\sigma$ and 3$\sigma$ intervals (1 \dof) for the three--flavour neutrino
  oscillation parameters from global data including solar, atmospheric,
  reactor (KamLAND and CHOOZ) and accelerator (K2K and MINOS) experiments.
}
\end{table}

Tab.~\ref{tab:summary2010} gives an updated summary of the present
best fit values and allowed ranges for the three-flavor oscillation
parameters.

\bibliographystyle{unsrt}

\end{document}